\documentclass[10pt,twocolumn]{article} 
\usepackage{simpleConference}
\usepackage{times}
\usepackage{graphicx}
\usepackage{amssymb}
\usepackage{booktabs}
\usepackage{url,hyperref}

\begin{document}

\newcommand{\BRpipeline}{Kraken~+~Ciaconna }
\newcommand{\ocrdmin}{Tesseract }
\newcommand{\ocrdvanilla}{Calamari GT4Hist }
\newcommand{\ocrdmax}{ocrdmax}

\title{Optical Character Recognition of 19\textsuperscript{th} Century Classical Commentaries: the Current State of Affairs\thanks{This version of the article corresponds to the accepted manuscript. For the final published version see \url{https://doi.org/10.1145/3476887.3476911}.}}

\author{Matteo Romanello\\
\\
Université de Lausanne\\
\\
matteo.romanello@unil.ch\\

\and 
Sven Najem-Meyer\\
\\
EPFL\\
\\
sven.najem-meyer@epfl.ch\\

\and 
Bruce Robertson\\
\\
Mount Allison University\\
\\
broberts@mta.ca\\
}

\maketitle
\thispagestyle{empty}

\begin{abstract}
Together with critical editions and translations, commentaries are one of the main genres of publication in literary and textual scholarship, and have a century-long tradition. Yet, the exploitation of thousands of digitized historical commentaries was hitherto hindered by the poor quality of Optical Character Recognition (OCR), especially on commentaries to Greek texts.
In this paper, we evaluate the performances of two pipelines suitable for the OCR of historical classical commentaries.
Our results show that \BRpipeline reaches a substantially lower character error rate (CER) than Tesseract/OCR-D on commentary sections with high density of polytonic Greek text (average CER 7\% vs. 13\%), while Tesseract/OCR-D is slightly more accurate than \BRpipeline on text sections written predominantly in Latin script (average CER 8.2\% vs. 8.4\%).
As part of this paper, we also release GT4HistComment, a small dataset with OCR ground truth for 19\textsuperscript{th} classical commentaries and Pogretra, a large collection of training data and pre-trained models for a wide variety of ancient Greek typefaces.
\end{abstract}

\section{Introduction}

\begin{figure*}[hbt!]
  \centering
  \includegraphics[width=\linewidth]{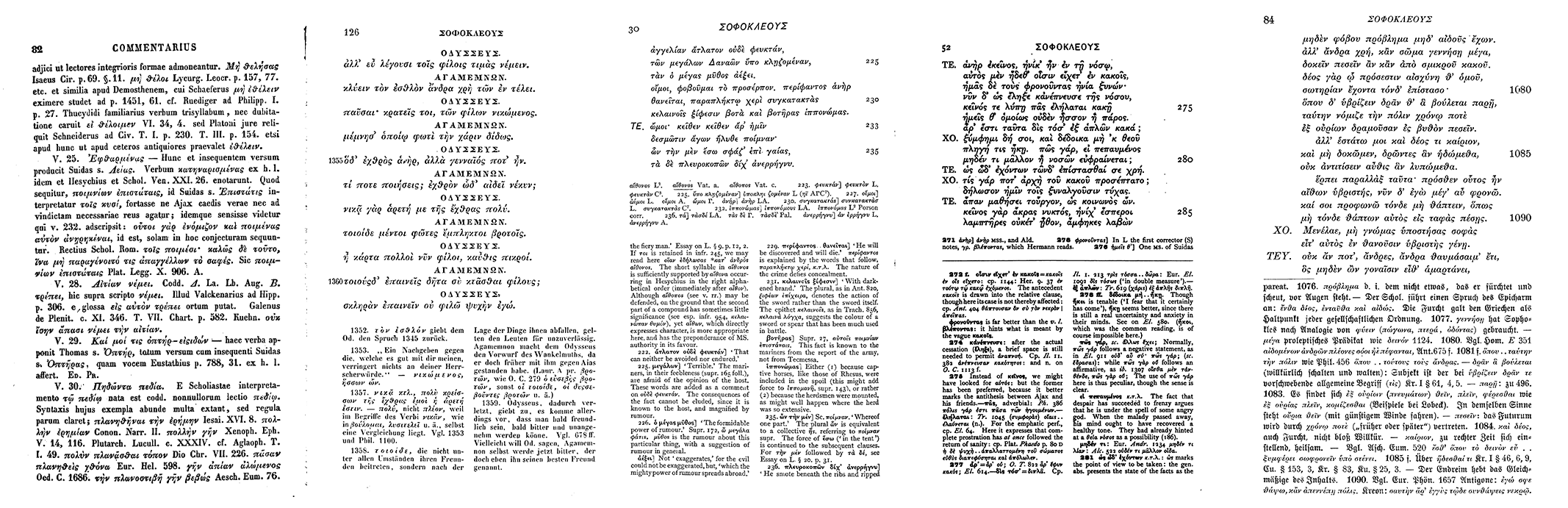}
  \caption{Example pages from 19\textsuperscript{th} century commentaries on Sophocles’ \textit{Ajax}. The commentaries are by (from left to right): Lobeck (1835), Schneidewin (1853), Campbell (1881), Jebb (1896) and Wecklein (1894).}
  \label{fig:commentaries-examples}
\end{figure*}

Recent large-scale digitization initiatives have greatly increased the access to cultural heritage collections. For researchers, however, their benefits remain highly dependent on the quality of their automatic transcription. Modern and contemporary historians, for example, are now in a relatively good position to exploit large collections of digitized newspapers, thanks to the reasonable accuracy reached by optical character recognition (OCR) systems on these materials (see e.g.~\cite{strobel_how_2020}). On the contrary, classicists have long been confronted to a paradoxical situation, where large amounts of digitized sources were openly available online, but where the poor performances of polytonic Greek OCR hampered their exploitation. This situation has started to change thanks to dramatic improvements on polytonic Greek OCR over the last decade. These advancements have enabled new ambitious projects such as the Free First Thousand Years of Greek (FF1KG)~\cite{muellner_free_2019-1}, an effort allied with the more extensive Open Greek \& Latin Project\footnote{\url{https://opengreekandlatin.org/}} (OGLP), which aims to create large open collections of digital editions of Greek and Latin primary sources. However, one type of historical document that has remained hitherto marginal---both to these efforts and their concomitant improvements---are classical commentaries. 

Scholarly commentaries are publications dedicated to the in-depth analysis and explanation of literary works, and are aimed at facilitating the reading and understanding of a given text. Moreover, they constitute a notable example of historical documents that present considerable challenges for OCR systems. However, because of the lack of ground truth data~\cite{smith_research_2018}, they are rarely considered by builders of OCR systems. Commentaries on Classical Greek and Latin literature present the typical difficulties of historical OCR; they have complex layouts, often with multiple columns and rows of text, and additionally contain Latin and Greek scripts mixed together within the same page and often even within the same line of text.
As a result, the OCR quality of commentaries available in open repositories such as the Internet Archive renders these materials inadequate, most of the times, for use in research contexts, despite their recognized potential~\cite{Heslin2016}.

In this paper we investigate the current state of affairs with regard to the OCR of historical Classical commentaries. We evaluate and compare the quality of OCR obtained with two open source OCR pipelines---OCR-D and OGLP's Ciaconna---on a new, multilingual ground truth dataset of 19\textsuperscript{th} century commentaries. We begin by presenting prior works related to the challenges of recognizing polytonic Greek and of historical OCR (Section \ref{sec:related-work}). 
In Section~\ref{sec:ajmc-project} we briefly present the Ajax Multi-Commentary project, which constitutes the context for the work presented in this paper, and in Section~\ref{GT-datasets} we describe the OCR ground truth datasets that we release as part of this work.
In Section~\ref{sec:experiments} we present the results of our experiments, and discuss the quality of OCR we achieved on these materials in the light of their intended future usage. We conclude the paper with some remarks on open challenges to be addressed by future work. 

\section{Related Work}
\label{sec:related-work}
Substantial efforts have been made to bring polytonic Greek OCR to a level of accuracy that would make its manual correction less costly than double-entry keying. Initial research in this area is characterized by the heavy use of OCR post-processing techniques, and had to deal with the lack of open source engines able to outperform ABBYY FineReader on texts with mixed Latin and polytonic Greek scripts, such as critical editions ~\cite{boschetti_improving_2009}. To tackle this issue, ground truth data was created~\cite{white_training_2012} to improve the ability of the open source engine Tesseract (version 2) to recognize polytonic Greek, and especially to deal with Greek diacritics (i.e. accents, breathing marks and their combinations). Another proposed solution~\cite{sichani_ocr_2019} consisted in using a k-nearest neighbors algorithm for character classification on top of the character segmentation and recognition performed by ABBYY, with substantial improvements in character recognition accuracy over ABBYY's output (94.2\% vs. 68.7\%).

From an OCR perspective, one of the main challenges of polytonic Greek OCR is the presence of diacritics, which results in a very large number of character classes to recognize (over 270). Furthermore, being very small, diacritics are often the source of errors. An approach to deal with this issue was proposed by Gatos et al. \cite{gatos_greek_2011} and consisted in combining multiple recognition modules, each of them dedicated to a specific class of characters or diacritics. 
Their OCR framework reached a character accuracy of 90\% and a word accuracy of 63\% on Greek texts. Another open source OCR software that was adapted to work on polytonic Greek is Gamera. A substantially extended version of Gamera became the basis for Rigaudon~\cite{robertson_large-scale_2017}, a processing pipeline developed for large scale OCR that employs image pre-processing and OCR post-processing to improve the accuracy of recognition, leading to an average character accuracy of 96\%. Finally, unlike the majority of previously discussed OCR systems, which require character segmentation, Katsouros et al. ~\cite{katsouros_recognition_2016} have developed a segmentation-free method that uses Hidden Markov Models (HMMs) to OCR text lines, reaching an average character accuracy of 92.39\% on the GRPOLY-DB dataset~\cite{gatos_grpoly-db_2015}.

More recently, deep neural architectures have proved to be very effective for OCR. Recurrent neural networks with Long Short Term Memory (LSTM), such as implemented in OCRopus 2\footnote{\url{https://github.com/ocropus/ocropy2}} and 3\footnote{\url{https://github.com/NVlabs/ocropus3}}, led to good results both on historical documents in English and German Fraktur script \cite{breuel_high-performance_2013}, and on historical polytonic Greek~\cite{simistira_recognition_2015}. As shown by Robertson \cite{OpticalCharacterRecognitionforClassicalPhilology}, these results can be improved even further with post-processing. This approach was implemented in Ciaconna\footnote{\url{https://github.com/brobertson/ciaconna}}, which applies spellcheckers and word dehyphenation to OCRopus' output. 

Moreover, hybrid neural architectures combining LSTMs with a convolutional layer were shown to improve LSTMs' results, for instance on early printed books (15\textsuperscript{th}-16\textsuperscript{th} century). Using Calamari\footnote{\url{https://github.com/Calamari-OCR}} \cite{wick_calamari_2020} with a hybrid architecture, Wick et al. \cite{wick_comparison_2018} report average improvements on CERs of 29.3\%, 34.9\% and 43.6\% when using respectively 60, 100 and 1,000 ground truth lines for training. Another open source implementation of this hybrid architecture is provided by Kraken~\cite{kiessling_kraken_2019}. On a sample of historical documents including polytonic Greek, the framework reached mean character accuracy rates ranging from 96.2\% to 99.5\% 
Finally, neural architectures for Handwriting Text Recognition (HTR) demonstrated their effectiveness also when applied to historical OCR. Ströbel et al. \cite{strobel_how_2020} trained the HTR+ model of the Transkribus framework~\cite{weidemann_htr_2017} and reached an average character recognition accuracy of 99.5\% on a historical German-language newspaper printed in black letter.


In addition to the development and improvement of OCR methods, recent efforts were devoted to the creation of platforms and frameworks aimed at facilitating working with OCR and at bringing the benefits of the latest advancements in OCR technologies to a larger public. Three main tools are worth mentioning in the context of historical Classical commentaries. First, Lace\footnote{\url{https://github.com/brobertson/Lace2}} is a GUI-based platform for the visualization and manual correction of OCR output, designed for automatic conversion to TEI-encoded XML. 
Second, OCR4all~\cite{reul_ocr4allopen-source_2019} is an open source OCR tool explicitly developed for users with no prior technical background, and especially those working on the earliest printed books. It implements an iterative workflow that allows for rapidly training very accurate OCR models for specific publications or publication series.
Lastly, OCR-D~\cite{engl_ocr-d_2020,engl_volltexte_2020} is a workflow-oriented, modular platform integrating several OCR engines into a common architecture. Unlike OCR4all, OCR-D was developed for a technical audience, such as staff working in the digitization units of cultural heritage institutions. Being an integrated platform, OCR-D is particularly suited for experimenting with various OCR tools and techniques, and exploring the effects of pre- and post-processing steps.\footnote{For detailed information about the OCR-D pipeline, we refer the reader to its extensive documentation available at https://ocr-d.de/ (in English and German).}
\section{The Ajax Multi-Commentary Project}
\label{sec:ajmc-project}


Scholarly commentaries do not only constitute reference works of absolute importance for any in-depth study of a text. They are also a fossilized testimony of reading and understanding a literary text as a historically-situated intellectual process. As such, commentaries can be used to study, from a history of science perspective, the history and evolution of this genre and of its exegetic practices. This is precisely the goal of Ajax Multi-Commentary project, which forms the context for the work presented in this paper.

The project focuses as a case study on the commentaries published about Sophocles’ \textit{Ajax}, a tragedy on the suicide of the Greek hero Ajax during the Trojan War. In order to enable an epistemological study of \textit{Ajax}'s commentaries, a digital multi-commentary is being developed in order to read, compare and analyze the entire commentary tradition of this tragedy. Layout detection and OCR of the digitized commentaries constitute the first necessary steps for the creation of multi-commentary interfaces. 
Further processing steps will entail the linking of the commentary section (or \textit{lemma}) with the portion of text it refers to, the alignment of commentaries with one another, as well as the semi-automatic enrichment of the commentary sections (e.g. by means of named entity processing and citation mining).   

\section{Ground Truth Datasets}
\label{GT-datasets}

While datasets such as GT4HistOCR~\cite{springmann_gt4histocr_2018, springmann_ground_2018} (Latin script) and GRPOLY-DB~\cite{gatos_grpoly-db_2015} (polytonic Greek script) do exist, there are no ground truth datasets with mixed Latin and polytonic Greek scripts that could be used for the evaluation of historical commentaries OCR. To start filling this gap, we release the OCR ground truth dataset for historical commentaries used for evaluation as well as the ground truth data for polytonic Greek that was used to train the Kraken models presented in Section \ref{sec:ogl-kraken-models}.

\subsection{OCR GT for Historical Commentaries}
\label{sect:commentaries-groundtruth}

GT4HistComment is a dataset of OCR ground truth for historical commentaries, created from the public domain subset of \textit{Ajax}'s commentaries.\footnote{\url{https://github.com/AjaxMultiCommentary/GT-commentaries-OCR}.} It consists of five 19\textsuperscript{th} century commentaries written in German, English, and Latin. Its main goal is to enable the evaluation of the OCR quality, especially in the light of further NLP processing that will be carried out during the Ajax Multi-Commentary project (see Section \ref{sec:ajmc-project}).

\begin{figure}[hbt!]
  \centering
  \includegraphics[width=\linewidth]{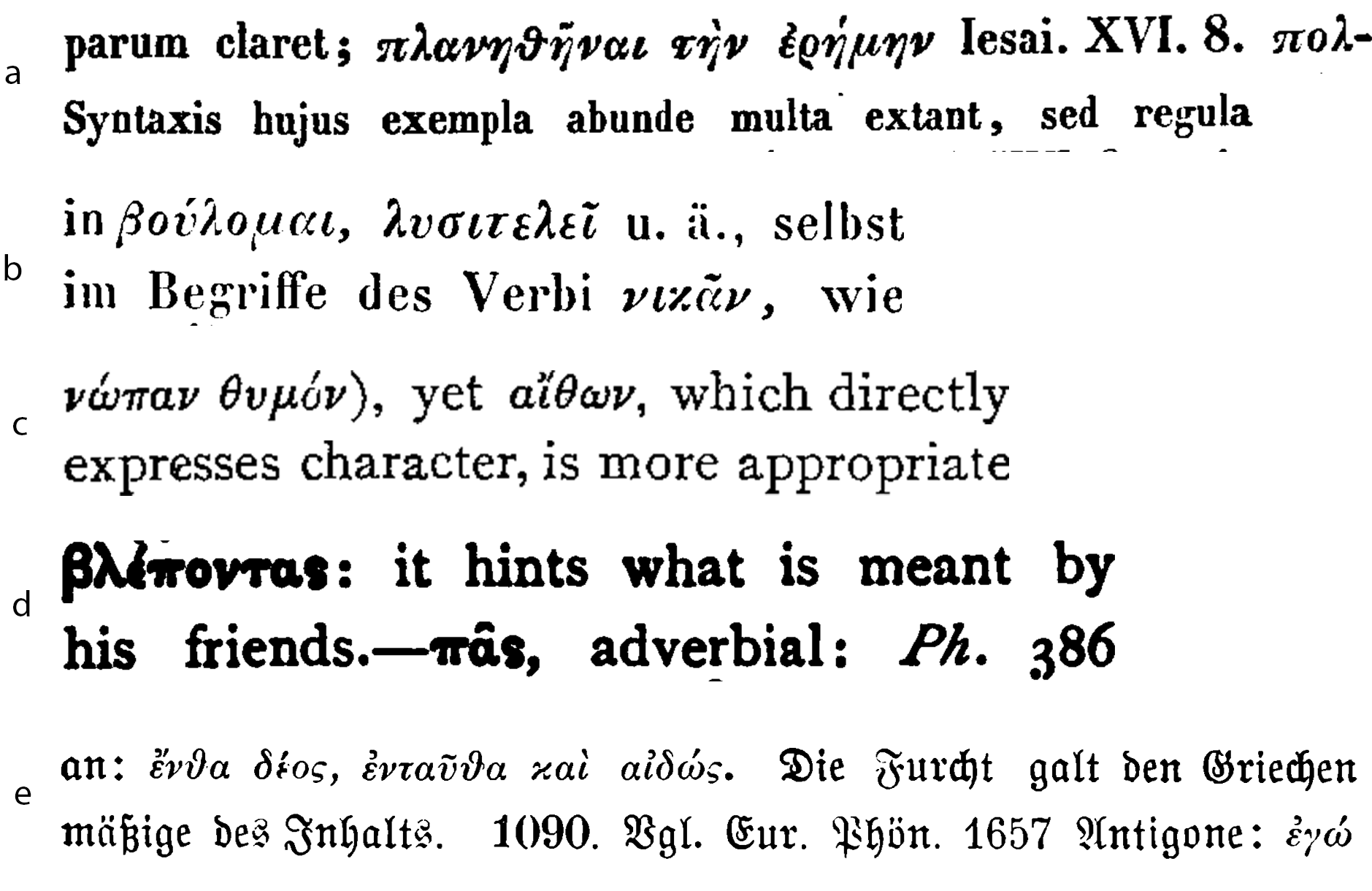}
  \caption{Example lines from selected commentaries on Sophocles' \textit{Ajax}: a) Lobeck, b) Schneidewin, c) Campbell, d) Jebb and e) Wecklein.}
  \label{fig:commentary-line-examples}
\end{figure}

The five commentaries in this dataset represent a variety of different layouts with varying degrees of complexity (see Fig. \ref{fig:commentaries-examples}). They feature multiple typefaces for the Latin and polytonic Greek scripts (Fig. \ref{fig:commentary-line-examples}).
The commentaries by Jebb (1896) and Campbell (1881) are written in English, Schneidewin (1853) and Wecklein (1894) in German, and Lobeck (1835) in Latin. All use serif fonts for Latin characters, with the exception of Wecklein, which is partly typeset in black letter. For Greek characters, the English commentaries use Porson (or Porson-like) fonts, whereas the others use a non-Teubner italicized font.


To create this dataset, the OCR output of the \BRpipeline pipeline (Section \ref{sec:ogl-kraken-models}) was manually corrected by a pool of annotators using the Lace platform. For the evaluation set, we selected a sample of five pages (2081±621 words) from each commentary.\footnote{ Detailed statistics (line, word and char-counts) can be found in the dataset's README.md file.} We deliberately decided against random sampling to ensure that this sample is representative of the different layout types (e.g. introduction, table of contents, commentary, etc.) and of their respective frequency within each commentary. Additionally to these, ten pages (2956±18 words) of ground truth were created only for selected commentaries and used to retrain the base \BRpipeline models (see Section~\ref{sec:base-vs-retrained-models}).

\begin{figure}[hbt!]
  \centering
  \includegraphics[width=\linewidth]{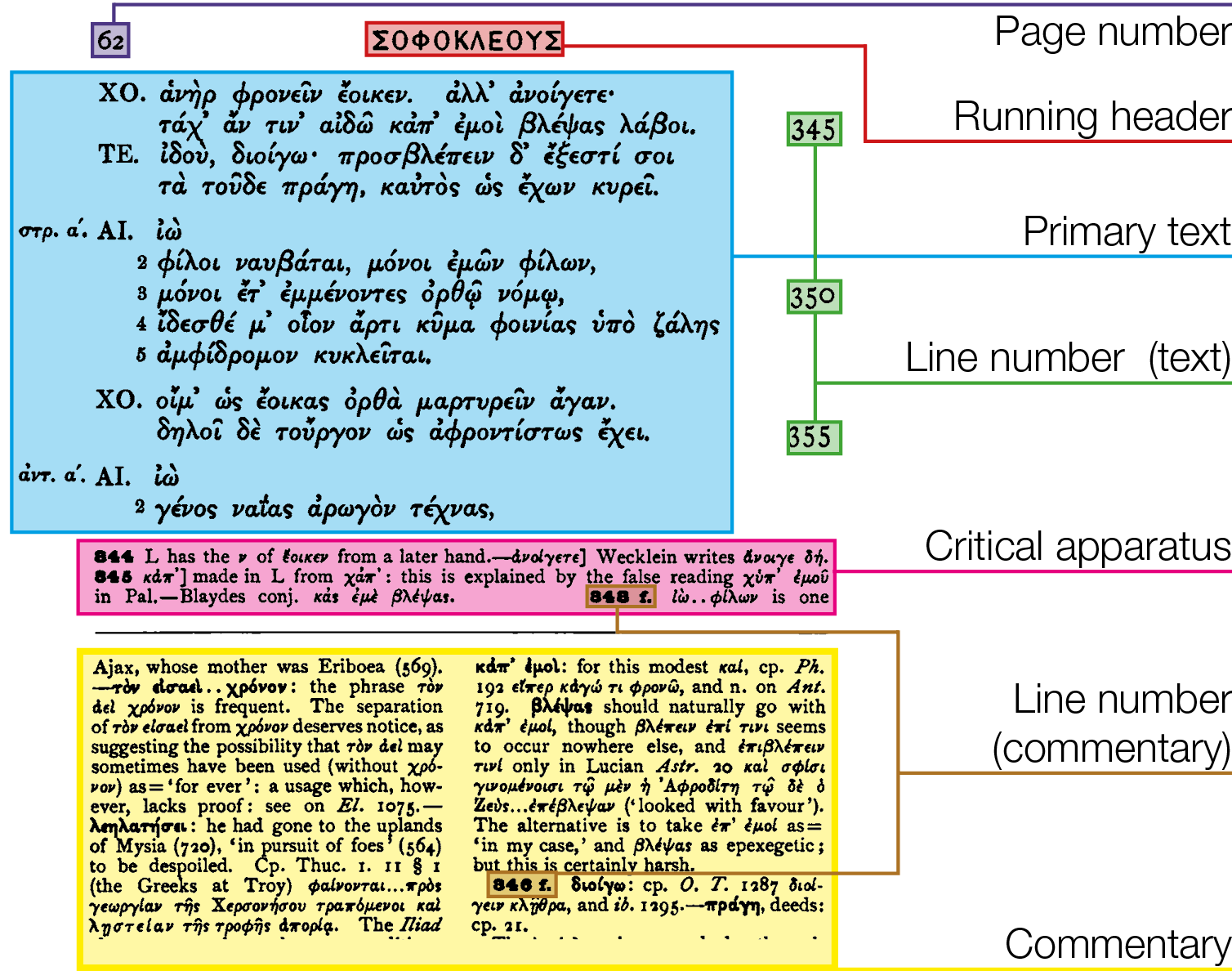}
  \caption{The main layout elements of a scholarly commentary page.}
  \label{fig:layout-regions}
\end{figure}

In order to ensure the quality of the ground truth datasets, an additional verification of all transcriptions made in Lace was carried out by an annotator on line-by-line pairs of image and corresponding text.  
Moreover, we annotated each page with OLR information; pages are divided into regions, and each region is assigned a type. OLR annotations were initially performed within Lace and subsequently refined with the help of the VGG Image Annotator (VIA)~\cite{dutta2019vgg}, according to a taxonomy that we devised specifically for commentaries. One of the usages of this fine-grained OLR annotation was to produce a detailed breakdown of character error rates according to region types (see Section \ref{sec:results-discussion}). 

While an in-depth discussion of our taxonomy goes beyond the scope of this paper, it is worth providing a brief presentation of the coarse-grained classes we defined (see Fig. \ref{fig:layout-regions}). These distinguish between the original Greek texts of the work being commented upon (\textit{Primary text}), the commentary sections (\textit{Commentary}), the commentator's translation of the commented text (\textit{Translation}), the section containing information about manuscript readings and editorial conjectures (\textit{Critical Apparatus}), the paratextual elements, such as table of contents, appendices, indices, footnotes, introductory and prefatory materials (\textit{Paratext}) and finally page numbers, line numbers alongside the primary text, as well as line numbers in the critical apparatus and commentary sections (\textit{Number}).

For evaluation purposes (see Table \ref{tab:cers-by-zone-types}), certain regions were grouped as follows : \emph{Commentary-like Texts} comprises commentaries and footnotes; \emph{Low Greek Texts}, comprises Latin script areas such as introduction, preface and translation; finally, \textit{Structured Texts} groups together appendices, bibliography, indices, titles and tables of contents.

\subsection{Polytonic Greek Training Data from Historic Texts (Pogretra)}
\label{sect:ogl-groundtruth}

Besides GT4HistComment, we release a substantial, carefully verified portion of the data produced by OGLP's editing process as the Pogretra dataset\footnote{Pogretra is published at \url{https://doi.org/10.5281/zenodo.4774200}.}, which constitutes the initial basis of the recognition of these commentaries in the \BRpipeline pipeline.

In addition to ground truth, this dataset includes several Kraken classifiers, each corresponding to a broad family of ancient Greek typefaces, as experiments to generate a Kraken classifier combining all historical typefaces did not yield satisfactory results.
 Of the two classifiers used in this study, the Porson classifier is well suited to typefaces common in the English-speaking world from the late 19\textsuperscript{th} century to present.
 Very different is the highly serifed family of Greek typefaces that became conventional in German-speaking countries of the 19\textsuperscript{th} century. Given this substantial difference, Pogretra provides a separate classifier for this typeface family, called `German-serifs'.  Because most pages of ancient Greek also comprise important Latin-script content (such as the critical apparatus), these classifiers recognize Latin characters as well. The Porson Kraken classifier used in this paper is trained on 6,607 lines of data comprising 79,823 words. The German-serifs Kraken classifier is trained on 25,365 lines of data and 218,806 words.

\section{Experiments} \label{sec:experiments}

\begin{table*}[h!]
\caption{Global and regional weight-averaged results (±std) of tested pipelines. CER scores ($0 \leq CER \leq 1$) are aggregated according to the following region groups (as explained in Section \ref{sect:commentaries-groundtruth}): Greek Texts (Greek), Commentary-like (Comm.), Low Greek Texts (Low Greek), Critical Apparatus (App. Crit.), Structured Texts (Struct.) and Numbers.}
\label{tab:cers-by-zone-types}
\centering
\resizebox{\textwidth}{!}{
\begin{tabular}{lccccccccc}
\toprule
Region & \multicolumn{3}{c}{Global} & Greek & Comm. &  Low-Greek &   App. Crit. & Struct. &  Numbers \\
Nb. of chars (\% Greek) & \multicolumn{3}{c}{51186 (29\%)} &   6657 (92\%) &     23825 (23\%) & 13322 (2\%) & 2062 (43\%) &       3371 (34\%) & 693 (0\%) \\
Metric &          F1 &      CER &      WER &          CER &             CER &        CER &        CER &              CER &      CER \\
\midrule
Calamari GT4Hist &     .61±.05 &   .30±.04 &  .39±.05 &      .84±.05 &         .27±.04 &    .05±.04 &    .46±.12 &          .34±.01 &  .23±.26 \\
Tesseract     &     \textbf{.82±.04} &  \textbf{.07±.02} &  \textbf{.16±.03} &      .13±.05 &         \textbf{.08±.02} &     \textbf{.01±.00} &    .12±.01 &          \textbf{.07±.01} &  \textbf{.13±.13} \\
Kraken+Ciaconna    &     .82±.02 &  .08±.02 &  .18±.02 &      \textbf{.07±.04} &         .11±.05 &    .04±.01 &     \textbf{.07±.00} &          .07±.02 &  .13±.17 \\
\bottomrule
\end{tabular}
}

\end{table*}


\begin{table*}[htbp!]
\caption{F1-Score, CER ($0 \leq CER \leq 1$) and normalized Levenstein distance (NLD) per commentary.}
\label{tab:CERs-by-commentary}
\resizebox{\textwidth}{!}{
\begin{tabular}{lrrrrrrrrrrrrrrr}
\toprule
Commentary & \multicolumn{3}{c}{Lobeck} & \multicolumn{3}{c}{Schneidewin} & \multicolumn{3}{c}{Campbell} & \multicolumn{3}{c}{Jebb} & \multicolumn{3}{c}{Wecklein} \\
Metric &     F1 &  CER &  NLD &          F1 &  CER &  NLD &       F1 &  CER &  NLD &   F1 &  CER &  NLD &       F1 &  CER &  NLD \\
\midrule
Calamari GT4Hist &   .52 & .37 & .63 &        .61 & .28 & .72 &     .67 & .27 & .73 & .63 & .31 & .69 &     .59 & .32 & .68 \\
Tesseract        &   .76 & .11 & .89 &        \textbf{.82} & \textbf{.08} & \textbf{.92} &     \textbf{.87} & \textbf{.05} & \textbf{.95} & \textbf{.80} & \textbf{.08} & \textbf{.92} &     .82 & .05 & .95 \\
Kraken+Ciaconna (base)  &   .75 & .11 & .89 &        .68 & .17 & .83 &     .83 & .07 & .93 &     .78 & .12 & .88 &     \textbf{.83} & \textbf{.05} & \textbf{.95} \\
Kraken+Ciaconna (retrained)  &   \textbf{.81} & \textbf{.09} & \textbf{.91} &        .82 & .09 & .91 &     - & - & - &    .82 & .09 & .91 &     - & - & - \\

\bottomrule
\end{tabular}
}

\end{table*}

In this section we present and discuss the evaluation results for two OCR pipelines: Tesseract/OCR-D (LSTM), and \BRpipeline (CNN-LSTM, same as Calamari). The Gamera software\footnote{\url{https://gamera.informatik.hsnr.de/}}, which provides support for polytonic Greek via the GreekOCR Toolkit add-on, was not considered as it proved less flexible than neural network-based pipelines due to its finite state machine-based architecture. It also requires a more laborious training process and its results are not superior.

\subsection{Comparing base and retrained models} \label{sec:base-vs-retrained-models}

We first investigated the benefits of adding 10 manually corrected pages to the baseline training material described in Section \ref{sect:ogl-groundtruth}. We performed this experiment on the three commentaries by Lobeck, Schneidewin and Jebb. The first two were selected as the respective base Kraken models were found to perform particularly poorly (Lace accuracy\footnote{We used the accuracy score provided by Lace as an initial indicator of OCR quality, which corresponds to the ratio of document tokens that are found in the dictionaries included in Lace.} <~60\%). Jebb, instead, was chosen as its training data could be reused also for Campbell's commentary, which uses identical typefaces. The improvement brought by retrained models varies substantially across commentaries. In fact, the CER of the retrained vs. base models for Lobeck, Schneidewin and Jebb decreases by 1.2\%, 8\% and 2.9\% respectively (see Table \ref{tab:CERs-by-commentary}).

\subsection{Comparing OCR pipelines}

\subsubsection{Tesseract/OCR-D}

Two series of experiments were conducted within the framework of OCR-D. First, we evaluated the performances of Calamari trained on GT4HistOCR \cite{springmann_gt4histocr_2018}. This model, however, was not trained on polytonic Greek data. Secondly, we tried two workflows using several default models from Tesseract in a confidence-based voting system. For each commentary, we took the model best suited for its language (English, Latin or German) and combined it with the default model for polytonic Greek. For both Schneidewin and Wecklein, we added a model trained on GT4HistOCR. Finally, for Wecklein only, we added three fraktur models to the confidence-based voting system\footnote{For the detailed list of Tesseract's models and training data, see \url{https://tesseract-ocr.github.io/tessdoc/\#traineddata-files-1}.}. The first workflow used Tesseract to segment and recognize binarized images. In the second workflow, images were segmented and pre-processed before being passed to Tesseract. However, since Tesseract could not be prevented from resegmenting words in its own fashion\footnote{See \url{https://github.com/OCR-D/ocrd-website/issues/215}.}, this workflow performed poorly (results not reported).

\subsubsection{Ciaconna}
\label{sec:ogl-kraken-models}
The Ciaconna OCR pipeline is based on Kraken and, occasionally, Tesseract. 
Training material is provided by the constantly expanding data from OGLP. As explained in Section \ref{sec:base-vs-retrained-models}, additional training data can be produced manually. Training uses NFC unicode normalization mode, which produces slightly better results, and data augmentation on scant data (less than 300 lines). To generate a classifier for Fraktur typefaces, we added the GT4HistOCR `RIDGES' material to Pogretra's german-serifs training set.\footnote{The GT4HistOCR collection is available at \url{https://zenodo.org/record/1344132}.}


In terms of OCR post-processing, Ciaconna applies dehyphenation and two kinds of polytonic Greek spellcheck to the OCR output. Dehyphenation aims to match every line ending in a hyphen with the first `word' of the next line, even if marginal line numbers intercede. When successful, the dehyphenated form is stored for later use. Ciaconna's first spellcheck uses a list of special Greek words which admit only one configuration of diacritics. Words from the output containing more than five character are then de-accented and matched with de-accented word-forms from the list. Successful matches are then correctly re-accented. The second spellcheck compares the OCR output with a list of 1.4 million unique Greek words. Ciaconna allows the replacement of up to one pair of commonly confused characters (or composed characters).


\subsection{Evaluation setting}

Due to the layout complexity of commentaries, different OCR-systems can output different reading orders of a same page, thus making it impractical to use a purely text-based approach for evaluation.
We therefore computed character error rates (CER) by using coordinates-based alignment of words, which allowed for a fast, region-based evaluation without need for carefully aligned text and line-images.\footnote{Combined diacritic-main forms (NFC unicode normalization mode) were used. Our OCR evaluator is available at \url{https://github.com/AjaxMultiCommentary/oclr}.}
For comparison with other research on historical OCR and on the impact of OCR quality on downstream NLP tasks, we report scores according to two additional metrics: PRImA TextEval's~\cite{clausner_flexible_2020} bag-of-words F1-score and normalized Levenshtein distance between ground truth and OCR output.

\subsection{Results and Discussion}
\label{sec:results-discussion}

The goal of the evaluation is twofold. Firstly, it aims to assess the strengths and weaknesses of each pipeline, both globally and regionally. Secondly, it seeks to determine whether the obtained OCR quality is acceptable for the type of analysis and further NLP steps planned in the AjMC project.

In Table \ref{tab:cers-by-zone-types} we report the results of the evaluation both at the global level and for groups of layout region types, showing the effect of polytonic Greek texts on the overall accuracy of each OCR pipeline. Unsurprisingly, an OCR pipeline without specific training for this script, such as \ocrdvanilla in our experiment, has high CERs in all sections that have a considerable proportion of Greek characters. While \ocrdmin and \BRpipeline perform similarly on text consisting of numbers or structured texts (e.g. indices, appendices, table of contents, etc.), the main differences in CERs are observed in text regions with a substantial presence of Greek characters. While \ocrdmin outperforms \BRpipeline in regions written mostly in Latin script, the latter remains more reliable in page regions such as primary text and critical apparatus, which are written predominantly in Greek script and/or are characterized by the presence of words that are usually not found in language models (e.g. people and place names, abbreviated Greek words).   

Moreover, the OCR quality deserves to be briefly discussed in the light of the upcoming NLP-steps to be performed on the OCRed texts. In fact, researchers have recently investigated the impact of OCR quality on downstream NLP tasks. Hill and Hengchen \cite{hill_quantifying_2019} found that texts with an F-score higher than 80\% can be reliably used for analysis such as topic modelling, collocations, vector space analysis and authorial attribution. van Strien et al. \cite{van_strien_assessing_2020} investigated the impact of OCR noise on sentence segmentation, named entity recognition and depedency parsing. They found that the least impact of OCR noise is observed on documents where the Normalized Levenshtein Distance (between ground truth and OCR output) is higher than .9. As reported in Table \ref{tab:CERs-by-commentary}, the OCR of all considered commentaries falls into the quality band where, according to both studies, further processing and analyses can be carried out with reliable results.  
\section{Conclusions and further work}

In this paper we evaluated the performances of Tesseract/OCR-D and \BRpipeline OCR pipelines on a dataset of 19\textsuperscript{th} century classical commentaries. We released also ground truth data and corresponding Kraken models produced by the Open Greek \& Latin project. It is hoped that these two datasets will spark further research on historical documents with mixed languages, and will lead to processing-ready versions of the thousands of commentaries available in the public domain. 

As it emerges from our evaluation results, Tesseract/OCR-D and \BRpipeline can be seen as two rather complementary pipelines with regards to the OCR of historical commentaries. The former performs better on texts written mostly in Latin script, while the latter excels on text with a large proportion of polytonic Greek, thanks also to extensive training data and the use of OCR post-processing. Based on these findings, we plan to explore how leveraging layout classification to perform OCR can lead to improved performances on this type of materials. In fact, classifying layout regions would allow for feeding different segments of the page to different OCR models and then combining these outputs in an ensemble-like system.

\bibliographystyle{abbrv}
\bibliography{ajmc-zotero}

\begin{thebibliography}{10}

\bibitem{boschetti_improving_2009}
F.~Boschetti, M.~Romanello, A.~Babeu, D.~Bamman, and G.~Crane.
\newblock Improving {{OCR Accuracy}} for {{Classical Critical Editions}}.
\newblock In M.~Agosti, J.~Borbinha, S.~Kapidakis, C.~Papatheodorou, and
  G.~Tsakonas, editors, {\em Research and {{Advanced Technology}} for {{Digital
  Libraries}}}, Lecture {{Notes}} in {{Computer Science}}, pages 156--167,
  {Berlin, Heidelberg}, 2009. {Springer}.

\bibitem{breuel_high-performance_2013}
T.~M. Breuel, A.~{Ul-Hasan}, M.~A. {Al-Azawi}, and F.~Shafait.
\newblock High-{{Performance OCR}} for {{Printed English}} and {{Fraktur Using
  LSTM Networks}}.
\newblock In {\em 2013 12th {{International Conference}} on {{Document
  Analysis}} and {{Recognition}}}, pages 683--687, {Washington, DC, USA}, Aug.
  2013. {IEEE}.

\bibitem{clausner_flexible_2020}
C.~Clausner, S.~Pletschacher, and A.~Antonacopoulos.
\newblock Flexible character accuracy measure for reading-order-independent
  evaluation.
\newblock {\em Pattern Recognition Letters}, 131:390--397, Mar. 2020.

\bibitem{dutta2019vgg}
A.~Dutta and A.~Zisserman.
\newblock The {{VIA}} annotation software for images, audio and video.
\newblock In {\em {{MM}} '19: {{Proceedings}} of the 27th {{ACM}} International
  Conference on Multimedia}, pages 2276--2279, {New York, NY, USA}, 2019.
  {ACM}.

\bibitem{engl_ocr-d_2020}
E.~Engl.
\newblock {OCR-D kompakt: Ergebnisse und Stand der Forschung in der
  F\"orderinitiative}.
\newblock {\em Bibliothek Forschung und Praxis}, 44(2):218--230, July 2020.

\bibitem{engl_volltexte_2020}
E.~Engl.
\newblock Volltexte f\"ur die {{Fr\"uhe Neuzeit}}.
\newblock {\em Zeitschrift f\"ur Historische Forschung}, 47(2):223--251, Apr.
  2020.

\bibitem{gatos_greek_2011}
B.~Gatos, G.~Louloudis, and N.~Stamatopoulos.
\newblock Greek {{Polytonic OCR Based}} on {{Efficient Character Class Number
  Reduction}}.
\newblock In {\em 2011 {{International Conference}} on {{Document Analysis}}
  and {{Recognition}}}, pages 1155--1159, {Beijing, China}, Sept. 2011. {IEEE}.

\bibitem{gatos_grpoly-db_2015}
B.~Gatos, N.~Stamatopoulos, G.~Louloudis, G.~Sfikas, G.~Retsinas,
  V.~Papavassiliou, F.~Sunistira, and V.~Katsouros.
\newblock {{GRPOLY}}-{{DB}}: {{An}} old {{Greek}} polytonic document image
  database.
\newblock In {\em 2015 13th {{International Conference}} on {{Document
  Analysis}} and {{Recognition}} ({{ICDAR}})}, pages 646--650, {Tunis,
  Tunisia}, Aug. 2015. {IEEE}.

\bibitem{Heslin2016}
P.~Heslin.
\newblock The dream of a universal variorum: {{Digitizing}} the commentary.
\newblock In C.~Kraus~Shuttleworth and {Christopher Stray}, editors, {\em
  Classical Commentaries : Explorations in a Scholarly Genre}, pages 494--511.
  {Oxford University Press}, {Oxford}, 2016.

\bibitem{hill_quantifying_2019}
M.~J. Hill and S.~Hengchen.
\newblock Quantifying the impact of dirty {{OCR}} on historical text analysis:
  {{Eighteenth Century Collections Online}} as a case study.
\newblock {\em Digital Scholarship in the Humanities}, 34(4):825--843, Dec.
  2019.

\bibitem{katsouros_recognition_2016}
V.~Katsouros, V.~Papavassiliou, F.~Simistira, and B.~Gatos.
\newblock Recognition of {{Greek Polytonic}} on {{Historical Degraded Texts
  Using HMMs}}.
\newblock In {\em 2016 12th {{IAPR Workshop}} on {{Document Analysis Systems}}
  ({{DAS}})}, pages 346--351, {Santorini, Greece}, Apr. 2016. {IEEE}.

\bibitem{kiessling_kraken_2019}
B.~Kiessling.
\newblock Kraken - an {{Universal Text Recognizer}} for the {{Humanities}}.
\newblock In {\em Digital {{Humanities}} 2019 {{Conference}} Papers}, {Utrecht,
  Netherlands}, 2019. {Utrecht University}.

\bibitem{muellner_free_2019-1}
L.~Muellner.
\newblock The {{Free First Thousand Years}} of {{Greek}}.
\newblock In M.~Berti, editor, {\em Digital {{Classical Philology}}}, chapter
  Digital Classical Philology, pages 7--18. {De Gruyter Saur}, {Berlin,
  Boston}, Aug. 2019.

\bibitem{reul_ocr4allopen-source_2019}
C.~Reul, D.~Christ, A.~Hartelt, N.~Balbach, M.~Wehner, U.~Springmann, C.~Wick,
  C.~Grundig, A.~B{\"u}ttner, and F.~Puppe.
\newblock {{OCR4all}}\textemdash{{An Open}}-{{Source Tool Providing}} a
  ({{Semi}}-){{Automatic OCR Workflow}} for {{Historical Printings}}.
\newblock {\em Applied Sciences}, 9:4853, Nov. 2019.

\bibitem{OpticalCharacterRecognitionforClassicalPhilology}
B.~Robertson.
\newblock Optical character recognition for classical philology.
\newblock In M.~Berti, editor, {\em Digital {{Classical Philology}}}, pages
  117--136. {De Gruyter Saur}, {Berlin, Boston}, Aug. 2019.

\bibitem{robertson_large-scale_2017}
B.~Robertson and F.~Boschetti.
\newblock Large-{{Scale Optical Character Recognition}} of {{Ancient Greek}}.
\newblock {\em Mouseion}, 14(3):341--359, Nov. 2017.

\bibitem{sichani_ocr_2019}
A.-M. Sichani, P.~Kaddas, G.~K. Mikros, and B.~Gatos.
\newblock {{OCR}} for {{Greek}} polytonic (multi accent) historical printed
  documents: Development, optimization and quality control.
\newblock In {\em Proceedings of the 3rd {{International Conference}} on
  {{Digital Access}} to {{Textual Cultural Heritage}}}, {{DATeCH2019}}, pages
  9--13, {New York, NY, USA}, May 2019. {Association for Computing Machinery}.

\bibitem{simistira_recognition_2015}
F.~Simistira, A.~{Ul-Hassan}, V.~Papavassiliou, B.~Gatos, V.~Katsouros, and
  M.~Liwicki.
\newblock Recognition of historical {{Greek}} polytonic scripts using {{LSTM}}
  networks.
\newblock In {\em 2015 13th {{International Conference}} on {{Document
  Analysis}} and {{Recognition}} ({{ICDAR}})}, pages 766--770, {Tunis,
  Tunisia}, Aug. 2015. {IEEE}.

\bibitem{smith_research_2018}
D.~A. Smith and R.~Cordell.
\newblock A {{Research Agenda}} for {{Historical}} and {{Multilingual Optical
  Character Recognition}}.
\newblock Technical report, {Northeastern University}, 2018.

\bibitem{springmann_ground_2018}
U.~Springmann, C.~Reul, S.~Dipper, and J.~Baiter.
\newblock Ground {{Truth}} for training {{OCR}} engines on historical documents
  in {{German Fraktur}} and {{Early Modern Latin}}.
\newblock {\em The Journal for Language Technology and Computational
  Linguistics}, 33(1):97--114, 2018.

\bibitem{springmann_gt4histocr_2018}
U.~Springmann, C.~Reul, S.~Dipper, and J.~Baiter.
\newblock {{GT4HistOCR}}: {{Ground Truth}} for training {{OCR}} engines on
  historical documents in {{German Fraktur}} and {{Early Modern Latin}}, Aug.
  2018.

\bibitem{strobel_how_2020}
P.~B. Str{\"o}bel, S.~Clematide, and M.~Volk.
\newblock How {{Much Data Do You Need}}? {{About}} the {{Creation}} of a
  {{Ground Truth}} for {{Black Letter}} and the {{Effectiveness}} of {{Neural
  OCR}}.
\newblock In {\em Proceedings of the 12th {{Language Resources}} and
  {{Evaluation Conference}}}, pages 3551--3559, {Marseille, France}, May 2020.
  {European Language Resources Association}.

\bibitem{van_strien_assessing_2020}
D.~{van Strien}, K.~Beelen, M.~Ardanuy, K.~Hosseini, B.~McGillivray, and
  G.~Colavizza.
\newblock Assessing the {{Impact}} of {{OCR Quality}} on {{Downstream NLP
  Tasks}}.
\newblock In {\em Proceedings of the 12th {{International Conference}} on
  {{Agents}} and {{Artificial Intelligence}}}, pages 484--496, {Valletta,
  Malta}, 2020. {SCITEPRESS - Science and Technology Publications}.

\bibitem{weidemann_htr_2017}
M.~Weidemann, J.~Michael, T.~Gr{\"u}ning, and R.~Labahn.
\newblock {{HTR Engine Based}} on {{NNs P2 Building}} deep architectures with
  {{TensorFlow}}.
\newblock Technical report, {READ}, Dec. 2017.

\bibitem{white_training_2012}
N.~White.
\newblock Training tesseract for ancient greek {{OCR}}.
\newblock {\em Eutypon}, (28-29):1--11, 2012.

\bibitem{wick_comparison_2018}
C.~Wick, C.~Reul, and F.~Puppe.
\newblock Comparison of {{OCR Accuracy}} on {{Early Printed Books}} using the
  {{Open Source Engines Calamari}} and {{OCRopus}}.
\newblock {\em Journal for Language Technology and Computational Linguistics
  (JLCL)}, 33(1):79--96, 2018.

\bibitem{wick_calamari_2020}
C.~Wick, C.~Reul, and F.~Puppe.
\newblock Calamari - {{A High}}-{{Performance Tensorflow}}-based {{Deep
  Learning Package}} for {{Optical Character Recognition}}.
\newblock {\em Digital Humanities Quarterly}, 014(2), June 2020.

\end{thebibliography}
\end{document}